\documentclass[aps,twocolumn,prl,superscriptaddress]{revtex4-1}

\usepackage{graphicx}
\usepackage{amsmath}
\usepackage{amssymb}
\usepackage{verbatim}
\usepackage{color}

\begin{document}
	
\title{Positional information readout in Ca$^{2+}$ signaling}

\author{Vaibhav H. Wasnik}
\affiliation{NCCR Chemical Biology, Departments of Biochemistry and Theoretical Physics, University of Geneva, 1211 Geneva, Switzerland}	
\affiliation{Indian Institute of Technology Goa, Ponda-403401, India}

\author{Peter Lipp}
\affiliation{Institute for Molecular Cell Biology, Research Centre for Molecular Imaging and Screening, Center for Molecular Signaling (PZMS), Medical Faculty, Saarland University, 66421 Homburg/Saar, Germany}

\author{Karsten Kruse}
\affiliation{NCCR Chemical Biology, Departments of Biochemistry and Theoretical Physics, University of Geneva, 1211 Geneva, Switzerland}

\date{\today}

\begin{abstract} 
Living cells respond to spatially confined signals. Intracellular signal transmission often involves the release of second messengers like Ca$^{2+}$. They eventually trigger a physiological response, for example, by activating kinases that in turn activate target proteins through phosphorylation. Here, we investigate theoretically how positional information can be accurately read out by protein phosphorylation in spite of rapid second messenger diffusion. We find that accuracy is increased by binding of kinases to the cell membrane prior to phosphorylation and by increasing the rate of Ca$^{2+}$ loss from the cell interior. These findings could explain some salient features of the conventional Protein Kinase C$\alpha$.
\end{abstract}

	
\maketitle

Living cells respond to external chemical and physical stimuli. In many cases, external factors result in global cellular responses with substrate-stiffness dependent cell differentiation being a particularly dramatic example~\cite{Engler:2006ga}. In other cases, signals carry spatial information on a subcellular scale~\cite{Rosse:2010gr}. In this way, localized uptake of extracellular material through endocytosis can be initiated~\cite{Godlee:2013fr} as well as targeted release through exocytosis~\cite{Oheim:2006jv}, amoeba migrate along chemical gradients~\cite{VanHaastert:2004bs}, neurons reinforce or weaken synapses~\cite{Deisseroth:1996ht,Wheeler:2008ja}, and immune cells polarize when making contact with antigen presenting cells~\cite{Kapsenberg:2003it}. 

Typically an external stimulus is translated into the release of a second messenger~\cite{Alberts2008}, for example, cyclic Adenosine-Monophosphate, Ca$^{2+}$ ions, and diacylglycerol (DAG). These then activate further downstream responses. For Ca$^{2+}$, this involves the Ca$^{2+}$ binding protein calmodulin (CaM) as well as the family of conventional Protein Kinases C (cPKCs). For activation, cPKC$\alpha$ requires simultaneous binding to DAG in the plasma membrane~\cite{Lipp:2011bh}. The signal is further relayed by phosphorylating target proteins, either directly as is the case for cPKC or indirectly by activating kinases as is the case for CaM. For example, the strength of synapses can be regulated by phosphorylating neuroreceptors and other synaptic proteins following a localized Ca$^{2+}$ release in the synapse~\cite{Groc:2004ds,Gerrow:2010kt}. The spatial distribution of phosphorylated proteins is thus a representation of the site of Ca$^{2+}$ release.  

Work on physical limits of detecting spatial information contained in cellular signals has so far focused on gradient sensing~\cite{Andrews:2007gu,Endres:2008eb,Hu:2010cl} and on extracting positional information from chemical gradients~\cite{Tostevin:2007hh}, for example, from the bicoid gradient in developing drosophila flies~\cite{Houchmandzadeh:2002uo,Gregor:2007du}. Also, a possible role of cell-cell communication for an efficient detection of shallow gradients has been investigated~\cite{Ellison:2016fd,Mugler:2016dy}. In this work, we ask how accurately cells can detect the position of a transient signal and consider the spatial distribution of phosphorylation events in response to localized Ca$^{2+}$ release. We find that kinases that are activated only after binding to the membrane detect the position of an incoming signal better than cytosolic kinases. Typically, the estimation error decreases with the rate at which Ca$^{2+}$ unbinds from the kinase and is lost from the system. Furthermore, it decreases more slowly than the inverse of the square root of the number of Ca$^{2+}$ ions in a signal.

\begin{figure}
\includegraphics[width=0.45\textwidth]{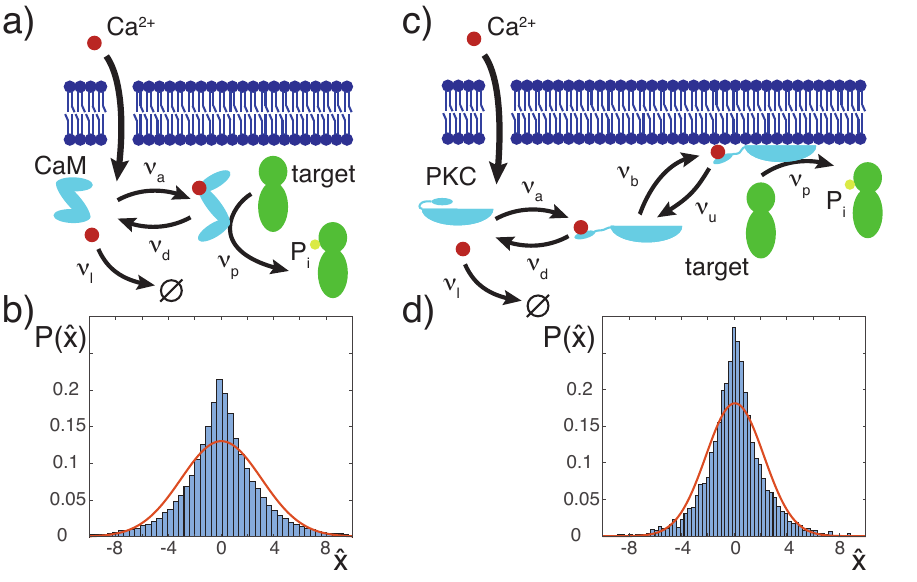}
\caption{\label{fig:scheme}Determination of the Ca$^{2+}$ entry site through phosphorylation of a target protein. a) Illustration of the CaM scenario.  Ca$^{2+}$ binds to a diffusible kinase at rate $\nu_a$, which then phosphorylates at rate $\nu_p$. Ca$^{2+}$ detaches at rate $\nu_d$ from the kinase and is lost from the system at rate $\nu_l$. Arrows indicate independent processes. b) Distribution of the estimated position $\hat x$ of Ca$^{2+}$ release given by averaging over the locations of the phosphorylation events and obtained from stochastic simulations. c) Illustration of the PKC scenario. The kinase binds to the membrane at rate $\nu_b$ and unbinds at rate $\nu_u$. Other parameters have the same meaning as in (a). d) Distribution of the estimated position $\hat x$ of Ca$^{2+}$ release for the PKC scenario obtained from stochastic simulations. Parameter values in (b) and (d) are $\nu_a/\nu_p=10$, $\nu_d/\nu_p=100$, $\nu_l/\nu_p=\nu_u/\nu_p=\nu_b/\nu_p=1$ and $D_K=0.01 D_C$. Space has been scaled with $\sqrt{D_C/\nu_p}$. Red lines in (b) and (d) indicate Gaussian fits to the distributions.}
\end{figure}
We start with the case of a diffusible kinase, which we assume to be abundant. In this case, different Ca$^{2+}$ ions are independent of each other as they do not compete for binding sites and we consider first a single Ca$^{2+}$ released into the cell interior at $x=0$, Fig.~\ref{fig:scheme}a. Below, we will use the results for a single Ca$^{2+}$ ion to treat the case of Ca$^{2+}$ puffs. We assume direct association of the Ca$^{2+}$ ion with the kinase at rate $\nu_a$. After binding Ca$^{2+}$, the kinase is active and phosphorylates target proteins at rate $\nu_p$. Ca$^{2+}$  dissociates from the kinase at rate $\nu_d$. Free Ca$^{2+}$ is lost from the system at rate $\nu_l$. The diffusion constants of Ca$^{2+}$ and the kinase are $D_C$ and $D_K$, respectively. Finally, we specify the geometry: the membrane is located at $z=0$ and extends infinitely into the $x$-direction. We neglect the dynamics in the $z$-direction and the intracellular space is the half-space with $z\ge0$. We will call this the calmodulin (CaM) scenario.  
		
For further analysis, we consider the case, in which the rate of target protein dephosphorylation is significantly lower than the overall rate at which a Ca$^{2+}$ ion leaves the system. In this way, all target proteins that have been phosphorylated as a consequence of Ca$^{2+}$ entry remain so at the time the ion is lost. We furthermore neglect any motion of the target proteins and are interested in the distribution of the phosphorylation events along the $x$-axis that have occurred before the Ca$^{2+}$ ion is lost. This amounts to averaging the response over time. We then consider the average position of the phosphorylation events along the $x$-axis, $\hat x$, as the estimated position of Ca$^{2+}$ release. 

In Figure~\ref{fig:scheme}b, we present the probability distribution $P$ of the estimated positions of  Ca$^{2+}$ release obtained from numerical simulations of 10$^6$ Ca$^{2+}$ release events. In our numerical simulations, we draw the time $\Delta t$ to the occurrence of the next event form an exponential distribution. The mean of this distribution is given by the inverse of the total rate of all reaction events possible in the present state (attachment and loss if Ca$^{2+}$ is not attached to the kinase, detachment and phosphorylation otherwise). We then draw the molecules' next position in $x$-direction from a Gaussian distribution with variance $2D_i\Delta t$, $i=C,K$. Then the actual event is determined and the corresponding action performed. The resulting distribution $P$ is centered around $x=0$ and more peaked than a Gaussian. 

Now consider a kinase that needs to bind to the membrane for activation, Fig.~\ref{fig:scheme}c. Membrane binding occurs at rate $\nu_b$ and unbinding at rate $\nu_u$. It has been shown that following Ca$^{2+}$ stimulation the translocation of cPKC to the membrane is independent of the cytoskeleton~\cite{Hui:2017jz}. Therefore, we focus our attention on diffusive transport of the kinase. On the membrane, diffusion is reduced compared to transport in the cytoplasm~\cite{LippincottSchwartz:2001dx}. For simplicity, we assume that a membrane-bound kinase is immobile. All other processes are the same as in the CaM scenario. We will refer to this case as the PKC scenario. 

In the numerical simulations of the PKC scenario, we have to account explicitly for the dynamics in the $z$-direction. In the simulation, the boundary is taken into account in the following way~\cite{Andrews:2004fs,Erban:2007fn}: if a diffusion step leads to a position outside the simulation domain, then the particle binds with a probability that is proportional to the binding rate $\nu_b$. In the opposite case, it is reflected. If the particle remains within the simulation domain after a diffusion step, there is still a possibility that it has bound to the membrane along its path. The corresponding probability is proportional to the binding rate and to a factor that depends on the distance of the particle to the domain boundary: $\exp\left\{-z(t)z(t')/(D_K(t'-t))\right\}$, where $z(t)$ and $z(t')$ are the $z$-coordinates of the particle at the time $t$ of the previous reaction event and the time $t'$ of the current reaction event. As for the CaM scenario, the distribution $P$ of estimated Ca$^{2+}$ release sites deviates from a Gaussian distribution, see Fig.~\ref{fig:scheme}d. Note, that for the same values of the phosphorylation, attachment, detachment, and loss rates, the distribution is narrower compared to the CaM scenario.

\begin{figure}
\includegraphics[width=0.45\textwidth]{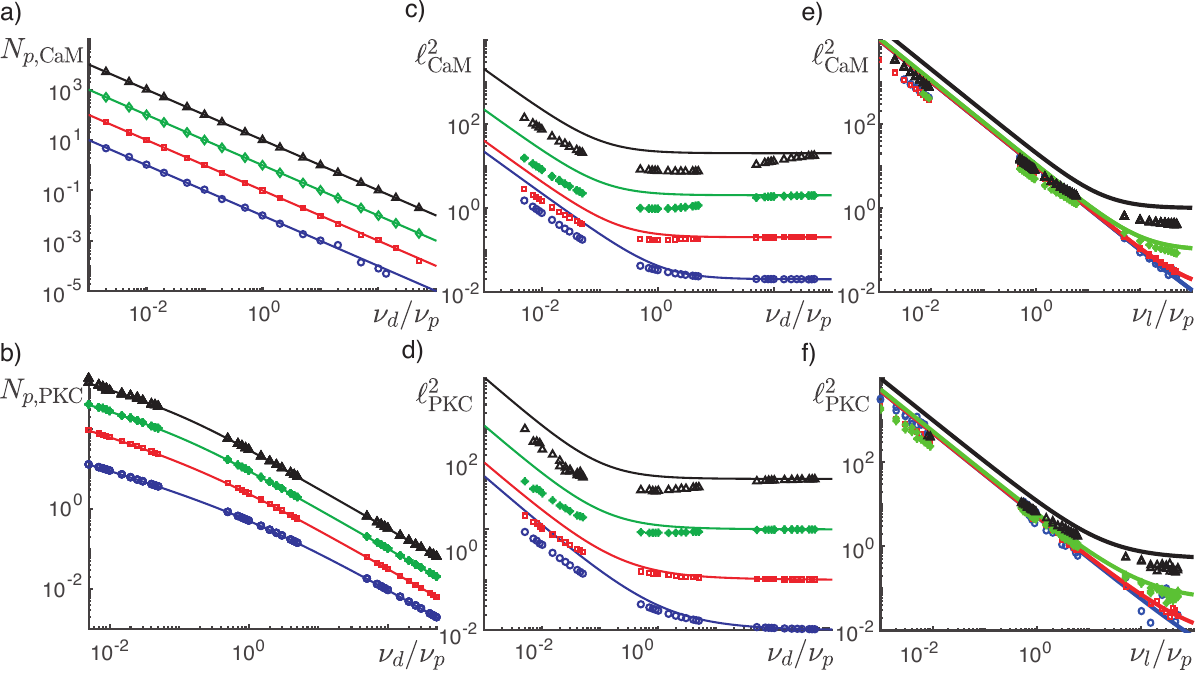}
\caption{\label{fig:parameterDependence}Parameter dependence of phosphorylation. a,b) Dependence of the average total number of phosphorylation events $\langle n\rangle\equiv N_p$ for the CaM- (a) and the PKC scenario (b). c-f) Dependence of the estimation error as a function of the detachment rate $\nu_d$ (c,d) and the loss rate $\nu_l$ (e,f) in the CaM- (c,e) and the PKC scenario (d,f). Symbols are for simulation results, lines are obtained from the mean-field calculations, see text. Parameter values are as in Fig.~\ref{fig:scheme} and $\nu_l/\nu_p=100$ ($\circ$, blue), $10$ ($\square$, red), $1$ ($\ast$, green), $0.1$ ($\triangle$, black) (a-d) and $\nu_d/\nu_p=100$ ($\circ$, blue), $10$ ($\square$, red), $1$ ($\ast$, green), $0.1$ ($\triangle$, black) (e,f). Space has been scaled with $\sqrt{D_C/\nu_p}$.}
\end{figure}

In both scenarios, the average total number $N_p$ of phosphorylation events is proportional to the phosphorylation rate and decreases with increasing detachment rate $\nu_d$, Fig.~\ref{fig:parameterDependence}a,b. In the CaM scenario, $N_\text{p,CaM}\propto\nu_d^{-1}$. In the PKC scenario we can observe two different scaling regimes as a function of $\nu_d$. Furthermore, $N_\text{p,PKC} \propto\nu_u^{-1}$.

Although the average number of phosphorylation events per Ca$^{2+}$ is an important characteristic of the signal detection process, it is not directly informative of the detection accuracy, which only depends on the (spatial) distribution of phosphorylation events. We define the error of the estimate to be given by the variance of the distribution $P$, $\ell^2=\int\text{d}\hat x\; {\hat x}^2P(\hat x)$. It decreases with increasing values of $\nu_d$ for $\nu_d\lesssim\nu_p$ and after a possible (weak) increase saturates, Fig.~\ref{fig:parameterDependence}c,d. For large enough detachment rates, the error is thus robust against changes in $\nu_d$. As a function of $\nu_l$ it decreases, Fig.~\ref{fig:parameterDependence}e,f. In the PKC scenario, the distribution of the estimated position is independent of the values of $\nu_b$ and $\nu_u$ as long as both are non-zero, because we assume membrane-bound particles to be immobile.
  
For a mean-field analysis of the above processes, let $p_C$ and $p_K$ be the respective probability distributions of free Ca$^{2+}$ and of the Ca$^{2+}$-kinase complex in the half space below the membrane. For the CaM scenario, we then have
\begin{align}
\label{eq:dCdt}
\partial_t p_C - D_C \Delta p_C  &=  \nu_d p_K - \nu_a p_C-\nu_{l}p_C\\
\label{eq:dKdt}
\partial_t p_K - D_K \Delta p_K  &= -\nu_d p_K + \nu_a p_C
\end{align}
with boundary conditions $\partial_z \left.p_C\right|_{z=0}=\partial_z\left. p_K\right|_{z=0}=0$. Under the mean-field assumption, the mean number $\hat n$ of phosphorylation events per unit length~\footnote{Note that $\hat n$ is not a probability distribution and typically $\int dx\;\hat n(x)\neq1$.} is given by
\begin{align}
\hat n(x) &= \nu_p  \int_0^\infty \text{d}z\int_0^\infty \text{d}t\; p_K(x,z,t)
\end{align}
in the limit $t\to\infty$. Using the initial conditions $p_K(x,z,t=0)=0$ and $p_C(x,z,t=0)=\delta(x)\delta(z)$, where $\delta$ is the Dirac $\delta$-distribution, we can integrate Eqs.~(\ref{eq:dCdt}) and (\ref{eq:dKdt}) with respect to $t$ from $0$ to $\infty$, solve them for $\int_0^\infty\text{d}t\; p_K(x,z,t)$, and finally obtain $\hat n$. The error is then 
\begin{align}
\ell^2_\text{CaM} &= \frac{ \int_{-\infty}^\infty\text{d}x\; x^2 \hat n(x)}{\int_{-\infty}^\infty\text{d}x\; \hat n(x)} = 2\left\{\ell_C^2 + \ell_K^2\left(1+\frac{\nu_a}{\nu_l} \right)\right\},
\label{eq:errorCalMF}
\end{align}
where $\ell_C^2\equiv D_C/\nu_l$ and $\ell_K^2\equiv D_K/\nu_d$ are the Ca$^{2+}$ and kinase diffusion-lengths, respectively. This expression agrees well with the simulation results, see Fig.~\ref{fig:parameterDependence}c,e. It is essentially given by the sum of the variances of Ca$^{2+}$ and kinase diffusion, where the latter is weighted by a factor depending on $\nu_a$ and $\nu_l$. The mean number of phosphorylation events by a single Ca$^{2+}$ ion, $N_\text{p,CaM}$ is
\begin{align} 
N_\text{p,CaM}&= \int_{-\infty}^\infty\text{d}x\; \hat n(x) = \frac{\nu_a \nu_{p}}{\nu_l \nu_{d}},
\label{eq:pnumberCalMF}
\end{align} 
which is equal to the exact expression for $N_\mathrm{p,CaM}$~[LongArticle].

Similarly, we can obtain $\hat n$ in the PKC scenario. In that case, the boundary condition on the kinase current in the $z$-direction is given by
\begin{align}
D_K\partial_z \left. p_K(x,z,t)\right|_{z=0}&=\nu_b p_K(x,z=0,t) - \nu_u p_k(x,t),
\end{align}
where $p_k$ is the distribution of PKC on the membrane. It is governed by
\begin{align}
\partial_t p_k(x,t) &=\nu_b p_K(x,z=0,t) -\nu_u p_k(x,t).
\end{align} 
The distribution of the mean number of phosphorylation events is now given by $\hat n=\nu_p \int_0^\infty \text{d}t\; p_k(x,t)$ and we find
\begin{align}
\ell^2_\text{PKC} &=  \frac{1}{2}\left[ \ell^2_\text{CaM} +\ell_C\ell_K\right]
\label{eq:errorPKCMF} \\ 
N_\text{p,PKC} & = \frac{\nu_b}{\nu_u} \left[2\ell^2_\text{PKC}+\ell_C\ell_K\right]^{-1/2}N_\text{p,CaM}.
\label{eq:pnumberPKCMF}
\end{align}
Note that in contrast to the CaM scenario, the mean number of phosphorylation events depends on the diffusion constants $D_C$ and $D_K$, because only kinases that make it to the membrane can phosphorylate. The mean-field result for the mean number of phosphorylation events by a single Ca$^{2+}$ ion, $N_\text{p,PKC}$, is exact~[LongArticle] and Eq.~(\ref{eq:errorPKCMF}) is a good approximation for the estimation error, see Fig.~\ref{fig:parameterDependence}b,d,f. Let us point out that $\ell^2_\text{PKC}<\ell^2_\text{CaM}$ for all parameter values, supporting that a membrane-binding kinase is better suited to detect the Ca$^{2+}$ entry point than a cytosolic kinase.

The differences between the membrane-binding and the cytosolic kinases as well as the dependence of the estimation error on the detachment and loss rates can be understood qualitatively. For a membrane-binding kinase, only Ca$^{2+}$ ions close to the membrane and thus typically also close to the Ca$^{2+}$ release site can contribute to target protein phosphorylation, because ions that are too far away will detach from the kinase before the latter binds to the membrane and is activated. In contrast, for a cytosolic kinase potentially all Ca$^{2+}$ ions can contribute. Increased detachment and loss rates $\nu_d$ and $\nu_l$ are expected to decrease the estimation error, because they reduce the time that a Ca$^{2+}$ ion can diffuse (bound to a kinase or not) before it phosphorylates. This is in agreement with the mean-field calculations and overall also with the stochastic simulations - only for $\nu_l\lesssim20$, the error increases slightly before saturating.

\begin{figure}
\includegraphics[width=0.45\textwidth]{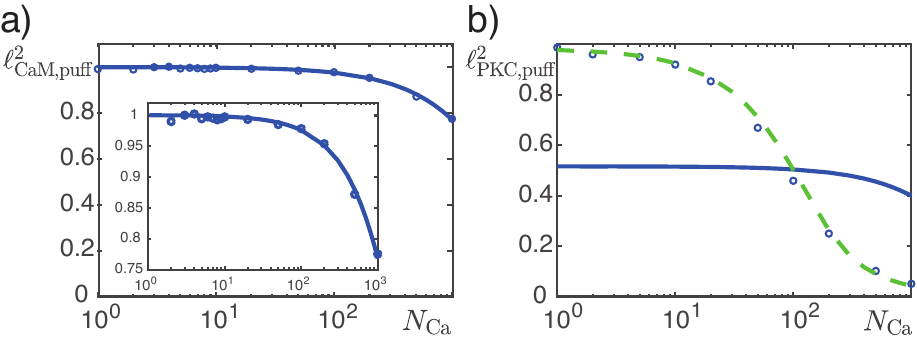}
\caption{\label{fig:puffs}Estimation error as a function of the number $N_\text{Ca}$ of Ca$^{2+}$ ions in a puff in the CaM- (a) and the PKC scenario (b). Inset in (a): different range of the error is shown. Circles indicate simulation results, full lines are from the mean-field calculation Eq.~(\ref{eq:variancePuff}), green dashed line in (b) is a fit of Eq.~(\ref{eq:variancePuff}) to the simulation data. Parameter values are $\nu_d/\nu_p=100$, $\nu_d/\nu_p=1$, $\nu_a/\nu_p=1$, and $\nu_l/\nu_p=10$. Other parameters as in Fig.~\ref{fig:scheme}. Space has been scaled with $\sqrt{D_C/\nu_p}$.}
\end{figure}

We now turn to Ca$^{2+}$ puffs. In Figure~\ref{fig:puffs}, we present the error as a function of the number $N_\text{Ca}$ of Ca$^{2+}$ in a puff. It does not decrease as $~1/N_\text{Ca}$ because not all Ca$^{2+}$ lead to a phosphorylation event. Note, that for $N_\text{Ca}\simeq1000$ the error in the PKC scenario is more than a factor 10 smaller compared to the CaM scenario. In both cases, the error starts to decrease as soon as $N_PN_\text{Ca}\approx1$.

We will now express the estimated error in the measurement performed by a puff through the distribution of phosphorylation events by one Ca$^{2+}$. Let $n(\xi)$ be the distribution of phosphorylation events resulting from a puff. A convenient notation for the variance $\ell_\text{puff}^2$ of the estimated position is in form of a path integral
\begin{align}
\ell_\text{puff}^2 &= \int \mathcal{D}n(\xi)\; {\hat \xi_{n(\xi)} }^2\mathcal{P}\left[n(\xi)\right],
\end{align}
where $\mathcal{P}$ is the probability distribution of the realizations and $\hat\xi_{n(\xi)}$ the estimated position for the distribution $n(\xi)$.

In the limit, where each Ca$^{2+}$ ion is resulting in phosphorylation at one position at most, phosphorylation at any two different positions results from two different Ca$^{2+}$ ions and are thus independent of each other. Consequently,
\begin{align}
\mathcal{P}\left[n(\xi)\right] &= \mathcal{N}\prod_\xi P(n,\xi)
\end{align}
with $P(n,\xi)$ being the probability of having $n$ phosphorylation events at $\xi$. We assume it to be given by a Poissonian distribution with a mean that is equal to the average phosphorylation profile $\hat n(\xi)$ of the distribution of phosphorylation events resulting from one Ca$^{2+}$ ion that was calculated above. Explicitly,
\begin{align}
P(n,\xi) & = \frac{\hat n(\xi)^n}{n!}\text{e}^{-\hat n(\xi)}.
\end{align}
After some calculation~[LongArticle], we find
\begin{align}
\ell_\text{puff}^2 &= \ell^2\frac{\mathrm{e}^{-N_\text{p}N_\text{Ca}}}{1-\mathrm{e}^{-N_\text{p}N_\text{Ca}}} \sum_{n=1}^\infty  \frac{N_\text{p}^nN_\text{Ca}^n}{n!n},
\label{eq:variancePuff}
\end{align}
where $N_\text{p}$ and $\ell^2$ are, respectively, the mean number of phosphorylation events and the variance of the corresponding distribution resulting from one Ca$^{2+}$ ion. For large $N_\text{Ca}$ we have $\ell^2_\text{puff}=\ell^2/(N_\text{p}N_\text{Ca})$. The mean-field expression is in good agreement with the simulation results in the CaM scenario and can be fitted to the data in case of the PKC scenario, see Fig.~\ref{fig:puffs}.

In conclusion, we have shown that the spatial distribution of phosphorylation events determines the site of Ca$^{2+}$ increase best when the Ca$^{2+}$ sensitive kinase requires membrane binding for activation. In this case, position estimation is optimized if the rate of Ca$^{2+}$ detachment from the kinase is comparable to the phosphorylation rate and if the rate of Ca$^{2+}$ loss from the system is maximal. We note that, for PKC$\alpha$, the Ca$^{2+}$ detachment rate is about 5 times that of the phosphorylation rate~\cite{Nalefski:2001wy}. Using experimental values for the various parameters, $D_C\approx500~\mu$m$^2$/s~\cite{Donahue:1987jx}, $D_K\approx10~\mu$m$^2$/s~\cite{Schaefer:2001fp}, $\nu_p\approx2$/s, $\nu_d\approx20$/s~\cite{Nalefski:2001wy}, and $\nu_l\approx40/s$~\cite{Smith:1998gg}, we find that the estimation error for PKC$\alpha$ and a single Ca$^{2+}$ ion is $\ell_\mathrm{PKC}^2\approx50\mu$m$^2$. This value decreases with increasing number of Ca$^{2+}$ ions in a puff.

In living cells there is always a background of Ca$^{2+}$ present, which can compromise the accuracy of the detection process of a localized stimulus. \textcolor{black}{In presence of background phosphorylation, the dependence of the error on parameters can change qualitatively. Notably, an increase of the Ca$^{2+}$ detachment rate $\nu_d$, which often leads to an increase of the accuracy in absence of background phosphorylation, Fig.~\ref{fig:parameterDependence}c,d, will lead to an increase of the error in its presence~[LongArticle]. The implementation of a threshold, such that only phosphorylation levels above the one induced by the background lead to a cell response, could at least partly remedy the detrimental effects of background phosphorylation.} %To avoid this effect, cells could implement a threshold such that only phosphorylation levels above the one induced by the background lead to a cell response. This could be implemented, for example, by reducing the average number of phosphorylation events per Ca$^{2+}$, by elevating the level of phosphorylation necessary for a response, or both. 
A full discussion of \textcolor{black}{the effects of background phosphorylation on reading out localized Ca$^{2+}$ signals} requires probably to consider a specific cell response.%, but see also Ref.~[LongArticle]. 

In future work, it will be interesting to consider aspects not accounted for in the present analysis. For example, cPKC$\alpha$ needs to bind to DAG for activation and forms clusters on the cell membrane~\cite{Bonny:2016em,Swanson:2016jj}. Also processes that are further "downstream" of target-protein phosphorylation like the diffusion of target proteins or actin-filament polymerization will affect the localization of the cell response. These studies should probably be restricted to specific processes, like the growth and maturation of a dendritic spine into a synapse. Our analysis, however, presents a general lower bound on the achievable accuracy.

\begin{acknowledgments}
We acknowledge funding through SFB 1027 by Deutsche Forschungsgemeinschaft. The computations were performed at University of Geneva on the Baobab cluster.
\end{acknowledgments}


\begin{thebibliography}{32}%
\makeatletter
\providecommand \@ifxundefined [1]{%
 \@ifx{#1\undefined}
}%
\providecommand \@ifnum [1]{%
 \ifnum #1\expandafter \@firstoftwo
 \else \expandafter \@secondoftwo
 \fi
}%
\providecommand \@ifx [1]{%
 \ifx #1\expandafter \@firstoftwo
 \else \expandafter \@secondoftwo
 \fi
}%
\providecommand \natexlab [1]{#1}%
\providecommand \enquote  [1]{``#1''}%
\providecommand \bibnamefont  [1]{#1}%
\providecommand \bibfnamefont [1]{#1}%
\providecommand \citenamefont [1]{#1}%
\providecommand \href@noop [0]{\@secondoftwo}%
\providecommand \href [0]{\begingroup \@sanitize@url \@href}%
\providecommand \@href[1]{\@@startlink{#1}\@@href}%
\providecommand \@@href[1]{\endgroup#1\@@endlink}%
\providecommand \@sanitize@url [0]{\catcode `\\12\catcode `\$12\catcode
  `\&12\catcode `\#12\catcode `\^12\catcode `\_12\catcode `\%12\relax}%
\providecommand \@@startlink[1]{}%
\providecommand \@@endlink[0]{}%
\providecommand \url  [0]{\begingroup\@sanitize@url \@url }%
\providecommand \@url [1]{\endgroup\@href {#1}{\urlprefix }}%
\providecommand \urlprefix  [0]{URL }%
\providecommand \Eprint [0]{\href }%
\providecommand \doibase [0]{http://dx.doi.org/}%
\providecommand \selectlanguage [0]{\@gobble}%
\providecommand \bibinfo  [0]{\@secondoftwo}%
\providecommand \bibfield  [0]{\@secondoftwo}%
\providecommand \translation [1]{[#1]}%
\providecommand \BibitemOpen [0]{}%
\providecommand \bibitemStop [0]{}%
\providecommand \bibitemNoStop [0]{.\EOS\space}%
\providecommand \EOS [0]{\spacefactor3000\relax}%
\providecommand \BibitemShut  [1]{\csname bibitem#1\endcsname}%
\let\auto@bib@innerbib\@empty
%</preamble>
\bibitem [{\citenamefont {Engler}\ \emph {et~al.}(2006)\citenamefont {Engler},
  \citenamefont {Sen}, \citenamefont {Sweeney},\ and\ \citenamefont
  {Discher}}]{Engler:2006ga}%
  \BibitemOpen
  \bibfield  {author} {\bibinfo {author} {\bibfnamefont {A.~J.}\ \bibnamefont
  {Engler}}, \bibinfo {author} {\bibfnamefont {S.}~\bibnamefont {Sen}},
  \bibinfo {author} {\bibfnamefont {H.~L.}\ \bibnamefont {Sweeney}}, \ and\
  \bibinfo {author} {\bibfnamefont {D.~E.}\ \bibnamefont {Discher}},\
  }\href@noop {} {\bibfield  {journal} {\bibinfo  {journal} {Cell}\ }\textbf
  {\bibinfo {volume} {126}},\ \bibinfo {pages} {677} (\bibinfo {year}
  {2006})}\BibitemShut {NoStop}%
\bibitem [{\citenamefont {Rosse}\ \emph {et~al.}(2010)\citenamefont {Rosse},
  \citenamefont {Linch}, \citenamefont {Kermorgant}, \citenamefont {Cameron},
  \citenamefont {Boeckeler},\ and\ \citenamefont {Parker}}]{Rosse:2010gr}%
  \BibitemOpen
  \bibfield  {author} {\bibinfo {author} {\bibfnamefont {C.}~\bibnamefont
  {Rosse}}, \bibinfo {author} {\bibfnamefont {M.}~\bibnamefont {Linch}},
  \bibinfo {author} {\bibfnamefont {S.}~\bibnamefont {Kermorgant}}, \bibinfo
  {author} {\bibfnamefont {A.~J.~M.}\ \bibnamefont {Cameron}}, \bibinfo
  {author} {\bibfnamefont {K.}~\bibnamefont {Boeckeler}}, \ and\ \bibinfo
  {author} {\bibfnamefont {P.~J.}\ \bibnamefont {Parker}},\ }\href@noop {}
  {\bibfield  {journal} {\bibinfo  {journal} {Nat.~Rev.~Mol.~Cell Bio.}\ }\textbf
  {\bibinfo {volume} {11}},\ \bibinfo {pages} {103} (\bibinfo {year}
  {2010})}\BibitemShut {NoStop}%
\bibitem [{\citenamefont {Godlee}\ and\ \citenamefont
  {Kaksonen}(2013)}]{Godlee:2013fr}%
  \BibitemOpen
  \bibfield  {author} {\bibinfo {author} {\bibfnamefont {C.}~\bibnamefont
  {Godlee}}\ and\ \bibinfo {author} {\bibfnamefont {M.}~\bibnamefont
  {Kaksonen}},\ }\href@noop {} {\bibfield  {journal} {\bibinfo  {journal} {J.
  Cell Biol.}\ }\textbf {\bibinfo {volume} {203}},\ \bibinfo {pages} {717}
  (\bibinfo {year} {2013})}\BibitemShut {NoStop}%
\bibitem [{\citenamefont {Oheim}\ \emph {et~al.}(2006)\citenamefont {Oheim},
  \citenamefont {Kirchhoff},\ and\ \citenamefont {St{\"u}hmer}}]{Oheim:2006jv}%
  \BibitemOpen
  \bibfield  {author} {\bibinfo {author} {\bibfnamefont {M.}~\bibnamefont
  {Oheim}}, \bibinfo {author} {\bibfnamefont {F.}~\bibnamefont {Kirchhoff}}, \
  and\ \bibinfo {author} {\bibfnamefont {W.}~\bibnamefont {St{\"u}hmer}},\
  }\href@noop {} {\bibfield  {journal} {\bibinfo  {journal} {Cell Calcium}\
  }\textbf {\bibinfo {volume} {40}},\ \bibinfo {pages} {423} (\bibinfo {year}
  {2006})}\BibitemShut {NoStop}%
\bibitem [{\citenamefont {Van~Haastert}\ and\ \citenamefont
  {Devreotes}(2004)}]{VanHaastert:2004bs}%
  \BibitemOpen
  \bibfield  {author} {\bibinfo {author} {\bibfnamefont {P.~J.~M.}\
  \bibnamefont {Van~Haastert}}\ and\ \bibinfo {author} {\bibfnamefont {P.~N.}\
  \bibnamefont {Devreotes}},\ }\href@noop {} {\bibfield  {journal} {\bibinfo
  {journal} {Nat.~Rev.~Mol.~Cell Bio.}\ }\textbf {\bibinfo {volume} {5}},\ \bibinfo
  {pages} {626} (\bibinfo {year} {2004})}\BibitemShut {NoStop}%
\bibitem [{\citenamefont {Deisseroth}\ \emph {et~al.}(1996)\citenamefont
  {Deisseroth}, \citenamefont {Bito},\ and\ \citenamefont
  {Tsien}}]{Deisseroth:1996ht}%
  \BibitemOpen
  \bibfield  {author} {\bibinfo {author} {\bibfnamefont {K.}~\bibnamefont
  {Deisseroth}}, \bibinfo {author} {\bibfnamefont {H.}~\bibnamefont {Bito}}, \
  and\ \bibinfo {author} {\bibfnamefont {R.~W.}\ \bibnamefont {Tsien}},\
  }\href@noop {} {\bibfield  {journal} {\bibinfo  {journal} {Neuron}\ }\textbf
  {\bibinfo {volume} {16}},\ \bibinfo {pages} {89} (\bibinfo {year}
  {1996})}\BibitemShut {NoStop}%
\bibitem [{\citenamefont {Wheeler}\ \emph {et~al.}(2008)\citenamefont
  {Wheeler}, \citenamefont {Barrett}, \citenamefont {Groth}, \citenamefont
  {Safa},\ and\ \citenamefont {Tsien}}]{Wheeler:2008ja}%
  \BibitemOpen
  \bibfield  {author} {\bibinfo {author} {\bibfnamefont {D.~G.}\ \bibnamefont
  {Wheeler}}, \bibinfo {author} {\bibfnamefont {C.~F.}\ \bibnamefont
  {Barrett}}, \bibinfo {author} {\bibfnamefont {R.~D.}\ \bibnamefont {Groth}},
  \bibinfo {author} {\bibfnamefont {P.}~\bibnamefont {Safa}}, \ and\ \bibinfo
  {author} {\bibfnamefont {R.~W.}\ \bibnamefont {Tsien}},\ }\href@noop {}
  {\bibfield  {journal} {\bibinfo  {journal} {J. Cell Biol.}\ }\textbf
  {\bibinfo {volume} {183}},\ \bibinfo {pages} {849} (\bibinfo {year}
  {2008})}\BibitemShut {NoStop}%
\bibitem [{\citenamefont {Kapsenberg}(2003)}]{Kapsenberg:2003it}%
  \BibitemOpen
  \bibfield  {author} {\bibinfo {author} {\bibfnamefont {M.~L.}\ \bibnamefont
  {Kapsenberg}},\ }\href@noop {} {\bibfield  {journal} {\bibinfo  {journal}
  {Nat.~Rev.~Immunol.}\ }\textbf {\bibinfo {volume} {3}},\ \bibinfo {pages} {984}
  (\bibinfo {year} {2003})}\BibitemShut {NoStop}%
\bibitem [{\citenamefont {Alberts}\ \emph {et~al.}(2008)\citenamefont
  {Alberts}, \citenamefont {Johnson}, \citenamefont {Lewis}, \citenamefont
  {Raff}, \citenamefont {Roberts},\ and\ \citenamefont {Walter}}]{Alberts2008}%
  \BibitemOpen
  \bibfield  {author} {\bibinfo {author} {\bibfnamefont {B.}~\bibnamefont
  {Alberts}}, \bibinfo {author} {\bibfnamefont {A.}~\bibnamefont {Johnson}},
  \bibinfo {author} {\bibfnamefont {J.}~\bibnamefont {Lewis}}, \bibinfo
  {author} {\bibfnamefont {M.}~\bibnamefont {Raff}}, \bibinfo {author}
  {\bibfnamefont {K.}~\bibnamefont {Roberts}}, \ and\ \bibinfo {author}
  {\bibfnamefont {P.}~\bibnamefont {Walter}},\ }\href@noop {} {\emph {\bibinfo
  {title} {{Molecular Biology of the Cell}}}},\ \bibinfo {edition} {5th}\ ed.,\
  edited by\ \bibinfo {editor} {\bibfnamefont {B.}~\bibnamefont {Alberts}}\
  (\bibinfo  {publisher} {Garland Science},\ \bibinfo {year}
  {2008})\BibitemShut {NoStop}%
\bibitem [{\citenamefont {Lipp}\ and\ \citenamefont
  {Reither}(2011)}]{Lipp:2011bh}%
  \BibitemOpen
  \bibfield  {author} {\bibinfo {author} {\bibfnamefont {P.}~\bibnamefont
  {Lipp}}\ and\ \bibinfo {author} {\bibfnamefont {G.}~\bibnamefont {Reither}},\
  }\href@noop {} {\bibfield  {journal} {\bibinfo  {journal} {Cold Spring Harb.~Perspect.~Biol.}\ }\textbf {\bibinfo {volume} {3}} (\bibinfo {year}
  {2011})}\BibitemShut {NoStop}%
\bibitem [{\citenamefont {Groc}\ \emph {et~al.}(2004)\citenamefont {Groc},
  \citenamefont {Heine}, \citenamefont {Cognet}, \citenamefont {Brickley},
  \citenamefont {Stephenson}, \citenamefont {Lounis},\ and\ \citenamefont
  {Choquet}}]{Groc:2004ds}%
  \BibitemOpen
  \bibfield  {author} {\bibinfo {author} {\bibfnamefont {L.}~\bibnamefont
  {Groc}}, \bibinfo {author} {\bibfnamefont {M.}~\bibnamefont {Heine}},
  \bibinfo {author} {\bibfnamefont {L.}~\bibnamefont {Cognet}}, \bibinfo
  {author} {\bibfnamefont {K.}~\bibnamefont {Brickley}}, \bibinfo {author}
  {\bibfnamefont {F.~A.}\ \bibnamefont {Stephenson}}, \bibinfo {author}
  {\bibfnamefont {B.}~\bibnamefont {Lounis}}, \ and\ \bibinfo {author}
  {\bibfnamefont {D.}~\bibnamefont {Choquet}},\ }\href@noop {} {\bibfield
  {journal} {\bibinfo  {journal} {Nat.~Neurosci.}\ }\textbf {\bibinfo {volume}
  {7}},\ \bibinfo {pages} {695} (\bibinfo {year} {2004})}\BibitemShut {NoStop}%
\bibitem [{\citenamefont {Gerrow}\ and\ \citenamefont
  {Triller}(2010)}]{Gerrow:2010kt}%
  \BibitemOpen
  \bibfield  {author} {\bibinfo {author} {\bibfnamefont {K.}~\bibnamefont
  {Gerrow}}\ and\ \bibinfo {author} {\bibfnamefont {A.}~\bibnamefont
  {Triller}},\ }\href@noop {} {\bibfield  {journal} {\bibinfo  {journal}
  {Curr.~Opin.~Neurobiol.}\ }\textbf {\bibinfo {volume} {20}},\
  \bibinfo {pages} {631} (\bibinfo {year} {2010})}\BibitemShut {NoStop}%
\bibitem [{\citenamefont {Andrews}\ and\ \citenamefont
  {Iglesias}(2007)}]{Andrews:2007gu}%
  \BibitemOpen
  \bibfield  {author} {\bibinfo {author} {\bibfnamefont {B.~W.}\ \bibnamefont
  {Andrews}}\ and\ \bibinfo {author} {\bibfnamefont {P.~A.}\ \bibnamefont
  {Iglesias}},\ }\href@noop {} {\bibfield  {journal} {\bibinfo  {journal} {PLoS
  Comput.~Biol.}\ }\textbf {\bibinfo {volume} {3}},\ \bibinfo {pages} {1489}
  (\bibinfo {year} {2007})}\BibitemShut {NoStop}%
\bibitem [{\citenamefont {Endres}\ and\ \citenamefont
  {Wingreen}(2008)}]{Endres:2008eb}%
  \BibitemOpen
  \bibfield  {author} {\bibinfo {author} {\bibfnamefont {R.~G.}\ \bibnamefont
  {Endres}}\ and\ \bibinfo {author} {\bibfnamefont {N.~S.}\ \bibnamefont
  {Wingreen}},\ }\href@noop {} {\bibfield  {journal} {\bibinfo  {journal} {Proc.~Natl.~Acad.~Sci.~USA}\ }\textbf {\bibinfo {volume} {105}},\ \bibinfo {pages}
  {15749} (\bibinfo {year} {2008})}\BibitemShut {NoStop}%
\bibitem [{\citenamefont {Hu}\ \emph {et~al.}(2010)\citenamefont {Hu},
  \citenamefont {Chen}, \citenamefont {Rappel},\ and\ \citenamefont
  {Levine}}]{Hu:2010cl}%
  \BibitemOpen
  \bibfield  {author} {\bibinfo {author} {\bibfnamefont {B.}~\bibnamefont
  {Hu}}, \bibinfo {author} {\bibfnamefont {W.}~\bibnamefont {Chen}}, \bibinfo
  {author} {\bibfnamefont {W.-J.}\ \bibnamefont {Rappel}}, \ and\ \bibinfo
  {author} {\bibfnamefont {H.}~\bibnamefont {Levine}},\ }\href@noop {}
  {\bibfield  {journal} {\bibinfo  {journal} {Phys.~Rev.~Lett.}\ }\textbf
  {\bibinfo {volume} {105}},\ \bibinfo {pages} {048104} (\bibinfo {year}
  {2010})}\BibitemShut {NoStop}%
\bibitem [{\citenamefont {Tostevin}\ \emph {et~al.}(2007)\citenamefont
  {Tostevin}, \citenamefont {ten Wolde},\ and\ \citenamefont
  {Howard}}]{Tostevin:2007hh}%
  \BibitemOpen
  \bibfield  {author} {\bibinfo {author} {\bibfnamefont {F.}~\bibnamefont
  {Tostevin}}, \bibinfo {author} {\bibfnamefont {P.~R.}\ \bibnamefont {ten
  Wolde}}, \ and\ \bibinfo {author} {\bibfnamefont {M.}~\bibnamefont
  {Howard}},\ }\href@noop {} {\bibfield  {journal} {\bibinfo  {journal} {PLoS
  Comput.~Biol.}\ }\textbf {\bibinfo {volume} {3}},\ \bibinfo {pages} {763}
  (\bibinfo {year} {2007})}\BibitemShut {NoStop}%
\bibitem [{\citenamefont {Houchmandzadeh}\ \emph {et~al.}(2002)\citenamefont
  {Houchmandzadeh}, \citenamefont {Wieschaus},\ and\ \citenamefont
  {Leibler}}]{Houchmandzadeh:2002uo}%
  \BibitemOpen
  \bibfield  {author} {\bibinfo {author} {\bibfnamefont {B.}~\bibnamefont
  {Houchmandzadeh}}, \bibinfo {author} {\bibfnamefont {E.}~\bibnamefont
  {Wieschaus}}, \ and\ \bibinfo {author} {\bibfnamefont {S.}~\bibnamefont
  {Leibler}},\ }\href@noop {} {\bibfield  {journal} {\bibinfo  {journal}
  {Nature}\ }\textbf {\bibinfo {volume} {415}},\ \bibinfo {pages} {798}
  (\bibinfo {year} {2002})}\BibitemShut {NoStop}%
\bibitem [{\citenamefont {Gregor}\ \emph {et~al.}(2007)\citenamefont {Gregor},
  \citenamefont {Tank}, \citenamefont {Wieschaus},\ and\ \citenamefont
  {Bialek}}]{Gregor:2007du}%
  \BibitemOpen
  \bibfield  {author} {\bibinfo {author} {\bibfnamefont {T.}~\bibnamefont
  {Gregor}}, \bibinfo {author} {\bibfnamefont {D.~W.}\ \bibnamefont {Tank}},
  \bibinfo {author} {\bibfnamefont {E.~F.}\ \bibnamefont {Wieschaus}}, \ and\
  \bibinfo {author} {\bibfnamefont {W.}~\bibnamefont {Bialek}},\ }\href@noop {}
  {\bibfield  {journal} {\bibinfo  {journal} {Cell}\ }\textbf {\bibinfo
  {volume} {130}},\ \bibinfo {pages} {153} (\bibinfo {year}
  {2007})}\BibitemShut {NoStop}%
\bibitem [{\citenamefont {Ellison}\ \emph {et~al.}(2016)\citenamefont
  {Ellison}, \citenamefont {Mugler}, \citenamefont {Brennan}, \citenamefont
  {Lee}, \citenamefont {Huebner}, \citenamefont {Shamir}, \citenamefont {Woo},
  \citenamefont {Kim}, \citenamefont {Amar}, \citenamefont {Nemenman},
  \citenamefont {Ewald},\ and\ \citenamefont {Levchenko}}]{Ellison:2016fd}%
  \BibitemOpen
  \bibfield  {author} {\bibinfo {author} {\bibfnamefont {D.}~\bibnamefont
  {Ellison}}, \bibinfo {author} {\bibfnamefont {A.}~\bibnamefont {Mugler}},
  \bibinfo {author} {\bibfnamefont {M.~D.}\ \bibnamefont {Brennan}}, \bibinfo
  {author} {\bibfnamefont {S.~H.}\ \bibnamefont {Lee}}, \bibinfo {author}
  {\bibfnamefont {R.~J.}\ \bibnamefont {Huebner}}, \bibinfo {author}
  {\bibfnamefont {E.~R.}\ \bibnamefont {Shamir}}, \bibinfo {author}
  {\bibfnamefont {L.~A.}\ \bibnamefont {Woo}}, \bibinfo {author} {\bibfnamefont
  {J.}~\bibnamefont {Kim}}, \bibinfo {author} {\bibfnamefont {P.}~\bibnamefont
  {Amar}}, \bibinfo {author} {\bibfnamefont {I.}~\bibnamefont {Nemenman}},
  \bibinfo {author} {\bibfnamefont {A.~J.}\ \bibnamefont {Ewald}}, \ and\
  \bibinfo {author} {\bibfnamefont {A.}~\bibnamefont {Levchenko}},\ }\href@noop
  {} {\bibfield  {journal} {\bibinfo  {journal} {Proc.~Natl.~Acad.~Sci.~USA}\
  }\textbf {\bibinfo {volume} {113}},\ \bibinfo {pages} {E679} (\bibinfo {year}
  {2016})}\BibitemShut {NoStop}%
\bibitem [{\citenamefont {Mugler}\ \emph {et~al.}(2016)\citenamefont {Mugler},
  \citenamefont {Levchenko},\ and\ \citenamefont {Nemenman}}]{Mugler:2016dy}%
  \BibitemOpen
  \bibfield  {author} {\bibinfo {author} {\bibfnamefont {A.}~\bibnamefont
  {Mugler}}, \bibinfo {author} {\bibfnamefont {A.}~\bibnamefont {Levchenko}}, \
  and\ \bibinfo {author} {\bibfnamefont {I.}~\bibnamefont {Nemenman}},\
  }\href@noop {} {\bibfield  {journal} {\bibinfo  {journal} {Proc.~Natl.~Acad.~Sci.~USA}\ }\textbf {\bibinfo {volume} {113}},\ \bibinfo {pages} {E689} (\bibinfo
  {year} {2016})}\BibitemShut {NoStop}%
\bibitem [{\citenamefont {Hui}\ \emph {et~al.}(2017)\citenamefont {Hui},
  \citenamefont {Sauer}, \citenamefont {Kaestner}, \citenamefont {Kruse},\ and\
  \citenamefont {Lipp}}]{Hui:2017jz}%
  \BibitemOpen
  \bibfield  {author} {\bibinfo {author} {\bibfnamefont {X.}~\bibnamefont
  {Hui}}, \bibinfo {author} {\bibfnamefont {B.}~\bibnamefont {Sauer}}, \bibinfo
  {author} {\bibfnamefont {L.}~\bibnamefont {Kaestner}}, \bibinfo {author}
  {\bibfnamefont {K.}~\bibnamefont {Kruse}}, \ and\ \bibinfo {author}
  {\bibfnamefont {P.}~\bibnamefont {Lipp}},\ }\href@noop {} {\bibfield
  {journal} {\bibinfo  {journal} {Sci.~Rep.}\ }\textbf {\bibinfo {volume} {7}}
  (\bibinfo {year} {2017})}\BibitemShut {NoStop}%
\bibitem [{\citenamefont {Lippincott-Schwartz}\ \emph
  {et~al.}(2001)\citenamefont {Lippincott-Schwartz}, \citenamefont {Snapp},\
  and\ \citenamefont {Kenworthy}}]{LippincottSchwartz:2001dx}%
  \BibitemOpen
  \bibfield  {author} {\bibinfo {author} {\bibfnamefont {J.}~\bibnamefont
  {Lippincott-Schwartz}}, \bibinfo {author} {\bibfnamefont {E.}~\bibnamefont
  {Snapp}}, \ and\ \bibinfo {author} {\bibfnamefont {A.}~\bibnamefont
  {Kenworthy}},\ }\href@noop {} {\bibfield  {journal} {\bibinfo  {journal} {Nat.~Rev.~Mol.~Cell Bio.}\ }\textbf {\bibinfo {volume} {2}},\ \bibinfo {pages} {444}
  (\bibinfo {year} {2001})}\BibitemShut {NoStop}%
\bibitem [{\citenamefont {Andrews}\ and\ \citenamefont
  {Bray}(2004)}]{Andrews:2004fs}%
  \BibitemOpen
  \bibfield  {author} {\bibinfo {author} {\bibfnamefont {S.~S.}\ \bibnamefont
  {Andrews}}\ and\ \bibinfo {author} {\bibfnamefont {D.}~\bibnamefont {Bray}},\
  }\href@noop {} {\bibfield  {journal} {\bibinfo  {journal} {Phys.~Biol.}\
  }\textbf {\bibinfo {volume} {1}},\ \bibinfo {pages} {137} (\bibinfo {year}
  {2004})}\BibitemShut {NoStop}%
\bibitem [{\citenamefont {Erban}\ and\ \citenamefont
  {Chapman}(2007)}]{Erban:2007fn}%
  \BibitemOpen
  \bibfield  {author} {\bibinfo {author} {\bibfnamefont {R.}~\bibnamefont
  {Erban}}\ and\ \bibinfo {author} {\bibfnamefont {S.~J.}\ \bibnamefont
  {Chapman}},\ }\href@noop {} {\bibfield  {journal} {\bibinfo  {journal}
  {Phys.~Biol.}\ }\textbf {\bibinfo {volume} {4}},\ \bibinfo {pages} {16}
  (\bibinfo {year} {2007})}\BibitemShut {NoStop}%
\bibitem [{Note1()}]{Note1}%
  \BibitemOpen
  \bibinfo {note} {Note that $p$ is not a probability distribution and
  typically $\DOTSI \intop \ilimits@ dx\protect \tmspace +\thickmuskip
  {.2777em}p(x)\not =1$.}\BibitemShut {Stop}%
\bibitem [{\citenamefont {Nalefski}\ and\ \citenamefont
  {Newton}(2001)}]{Nalefski:2001wy}%
  \BibitemOpen
  \bibfield  {author} {\bibinfo {author} {\bibfnamefont {E.~A.}\ \bibnamefont
  {Nalefski}}\ and\ \bibinfo {author} {\bibfnamefont {A.~C.}\ \bibnamefont
  {Newton}},\ }\href@noop {} {\bibfield  {journal} {\bibinfo  {journal}
  {Biochemistry}\ }\textbf {\bibinfo {volume} {40}},\ \bibinfo {pages} {13216}
  (\bibinfo {year} {2001})}\BibitemShut {NoStop}%
\bibitem [{\citenamefont {Donahue}\ and\ \citenamefont
  {Abercrombie}(1987)}]{Donahue:1987jx}%
  \BibitemOpen
  \bibfield  {author} {\bibinfo {author} {\bibfnamefont {B.~S.}\ \bibnamefont
  {Donahue}}\ and\ \bibinfo {author} {\bibfnamefont {R.~F.}\ \bibnamefont
  {Abercrombie}},\ }\href@noop {} {\bibfield  {journal} {\bibinfo  {journal}
  {Cell Calcium}\ }\textbf {\bibinfo {volume} {8}},\ \bibinfo {pages} {437}
  (\bibinfo {year} {1987})}\BibitemShut {NoStop}%
\bibitem [{\citenamefont {Schaefer}\ \emph {et~al.}(2001)\citenamefont
  {Schaefer}, \citenamefont {Albrecht}, \citenamefont {Hofmann}, \citenamefont
  {Gudermann},\ and\ \citenamefont {Schultz}}]{Schaefer:2001fp}%
  \BibitemOpen
  \bibfield  {author} {\bibinfo {author} {\bibfnamefont {M.}~\bibnamefont
  {Schaefer}}, \bibinfo {author} {\bibfnamefont {N.}~\bibnamefont {Albrecht}},
  \bibinfo {author} {\bibfnamefont {T.}~\bibnamefont {Hofmann}}, \bibinfo
  {author} {\bibfnamefont {T.}~\bibnamefont {Gudermann}}, \ and\ \bibinfo
  {author} {\bibfnamefont {G.}~\bibnamefont {Schultz}},\ }\href@noop {}
  {\bibfield  {journal} {\bibinfo  {journal} {FASEB J.}\ }\textbf
  {\bibinfo {volume} {15}},\ \bibinfo {pages} {1634} (\bibinfo {year}
  {2001})}\BibitemShut {NoStop}%
\bibitem [{\citenamefont {Smith}\ \emph {et~al.}(1998)\citenamefont {Smith},
  \citenamefont {Keizer}, \citenamefont {Stern}, \citenamefont {Lederer},
  \citenamefont {journal},\ and\ \citenamefont {{1998}}}]{Smith:1998gg}%
  \BibitemOpen
  \bibfield  {author} {\bibinfo {author} {\bibfnamefont {G.~D.}\ \bibnamefont
  {Smith}}, \bibinfo {author} {\bibfnamefont {J.~E.}\ \bibnamefont {Keizer}},
  \bibinfo {author} {\bibfnamefont {M.~D.}\ \bibnamefont {Stern}}, \bibinfo
  {author} {\bibfnamefont {W.~J.}\ \bibnamefont {Lederer}}, \ and\ \bibinfo {author}
  {\bibfnamefont {H.}\ \bibnamefont {Cheng}},\ }\href@noop {} {\bibfield  {journal} {\bibinfo
  {journal} {Biophys.~J.}\ }\textbf {\bibinfo {volume} {75}},\ \bibinfo {pages}
  {15} (\bibinfo {year} {1998})}\BibitemShut {NoStop}%
\bibitem [{\citenamefont {Bonny}\ \emph {et~al.}(2016)\citenamefont {Bonny},
  \citenamefont {Hui}, \citenamefont {Schweizer}, \citenamefont {Kaestner},
  \citenamefont {Zeug}, \citenamefont {Kruse},\ and\ \citenamefont
  {Lipp}}]{Bonny:2016em}%
  \BibitemOpen
  \bibfield  {author} {\bibinfo {author} {\bibfnamefont {M.}~\bibnamefont
  {Bonny}}, \bibinfo {author} {\bibfnamefont {X.}~\bibnamefont {Hui}}, \bibinfo
  {author} {\bibfnamefont {J.}~\bibnamefont {Schweizer}}, \bibinfo {author}
  {\bibfnamefont {L.}~\bibnamefont {Kaestner}}, \bibinfo {author}
  {\bibfnamefont {A.}~\bibnamefont {Zeug}}, \bibinfo {author} {\bibfnamefont
  {K.}~\bibnamefont {Kruse}}, \ and\ \bibinfo {author} {\bibfnamefont
  {P.}~\bibnamefont {Lipp}},\ }\href@noop {} {\bibfield  {journal} {\bibinfo
  {journal} {Sci.~Rep.}\ }\textbf {\bibinfo {volume} {6}},\ \bibinfo {pages}
  {36028} (\bibinfo {year} {2016})}\BibitemShut {NoStop}%
\bibitem [{\citenamefont {Swanson}\ \emph {et~al.}(2016)\citenamefont
  {Swanson}, \citenamefont {Sommese}, \citenamefont {Petersen}, \citenamefont
  {Ritt}, \citenamefont {Karslake}, \citenamefont {Thomas},\ and\ \citenamefont
  {Sivaramakrishnan}}]{Swanson:2016jj}%
  \BibitemOpen
  \bibfield  {author} {\bibinfo {author} {\bibfnamefont {C.~J.}\ \bibnamefont
  {Swanson}}, \bibinfo {author} {\bibfnamefont {R.~F.}\ \bibnamefont
  {Sommese}}, \bibinfo {author} {\bibfnamefont {K.~J.}\ \bibnamefont
  {Petersen}}, \bibinfo {author} {\bibfnamefont {M.}~\bibnamefont {Ritt}},
  \bibinfo {author} {\bibfnamefont {J.}~\bibnamefont {Karslake}}, \bibinfo
  {author} {\bibfnamefont {D.~D.}\ \bibnamefont {Thomas}}, \ and\ \bibinfo
  {author} {\bibfnamefont {S.}~\bibnamefont {Sivaramakrishnan}},\ }\href@noop
  {} {\bibfield  {journal} {\bibinfo  {journal} {PLoS ONE}\ }\textbf {\bibinfo
  {volume} {11}},\ \bibinfo {pages} {e0162331} (\bibinfo {year}
  {2016})}\BibitemShut {NoStop}%
\end{thebibliography}
\end{document}